\documentclass[conference]{IEEEtran}
\IEEEoverridecommandlockouts

\usepackage{graphicx}

\graphicspath{{figures/}{pictures/}{images/}{./}{../figures/}} 

\usepackage{cite}
\usepackage{amsmath,amssymb,amsfonts}
\usepackage{amsthm}
\usepackage{algorithmic}
\usepackage{booktabs}
\usepackage{enumitem}
\usepackage{graphicx}
\usepackage{textcomp}
\usepackage[dvipsnames]{xcolor}
\usepackage{subcaption}
\usepackage[frozencache,cachedir=.]{minted}
\usepackage{multirow}
\usepackage{siunitx}
\usepackage{tikz}
\usetikzlibrary{arrows.meta}
\usepackage{listings}
\usepackage[normalem]{ulem}

\usepackage{url}

\usepackage{breakurl}

\usepackage{hyperref}
\hypersetup{
    bookmarks=true,         
    unicode=false,          
    pdftoolbar=true,        
    pdfmenubar=true,        
    pdffitwindow=false,     
    pdfstartview={FitH},    
    pdftitle={cozy: Comparative Symbolic Execution for Binary  Programs},    
    pdfauthor={Caleb Helbling, Graham Leach-Krouse, Sam Lasser, and Greg Sullivan},     
    pdfsubject={cozy},   
    pdfcreator={Caleb Helbling},   
    pdfproducer={}, 
    pdfkeywords={symbolic execution, binary analysis, visualization}, 
    pdfnewwindow=true,      
    colorlinks=true,       
    linkcolor=blue,          
    citecolor=blue,        
    filecolor=blue,         
    urlcolor=blue        
}

\newif\ifdraft\draftfalse 

\newcommand{\FINISH}[3]{}
\newcommand{\sml}[1]{\FINISH{red}{SML}{#1}}

\newcommand*\circled[1]{\tikz[baseline=(char.base)]{
            \node[shape=circle,draw,inner sep=1pt] (char) {#1};}}

\newcommand{\toolname}{\texttt{cozy}}
\newcommand{\angrname}{\texttt{angr}}

\newcounter{nalg}[section] 
\lstnewenvironment{algorithm}[1][] 
{   
    \refstepcounter{nalg} 
    \lstset{ 
        mathescape=true,
        frame=tB,
        numbers=left, 
        numberstyle=\tiny,
        basicstyle=\scriptsize, 
        columns=fullflexible,
        keywordstyle=\color{black}\bfseries\em,
        keywords={,input, output, return, datatype, function, in, if, else, foreach, while, begin, end, } 
        numbers=left,
        xleftmargin=12pt
        #1 
    }
}
{}

\newtheorem{lemma}{Lemma}

\newtheorem{definition}{Definition}

\def\BibTeX{{\rm B\kern-.05em{\sc i\kern-.025em b}\kern-.08em
    T\kern-.1667em\lower.7ex\hbox{E}\kern-.125emX}}

\ifCLASSINFOpdf
\else
\fi
\begin{document}
%
\title{\toolname{}: Comparative Symbolic Execution for Binary Programs
  \thanks{This material is based upon work supported by the Defense Advanced Research Projects Agency (DARPA) and the Naval Information Warfare Center (NIWC) Pacific, under Contract No. N66001-20-C-4018. The views, opinions and/or findings expressed are those of the author and should not be interpreted as representing the official views or policies of the Department of Defense or the U.S. Government. Distribution Statement ``A'' (Approved for Public Release, Distribution Unlimited).}}


\author{\IEEEauthorblockN{Caleb Helbling, Graham Leach-Krouse, Sam Lasser, and Greg Sullivan}
  \IEEEauthorblockA{
    Draper \\
    \{chelbling, gleach-krouse, slasser, gsullivan\}@draper.com}
}
	

%


\IEEEoverridecommandlockouts
\makeatletter\def\@IEEEpubidpullup{2.5\baselineskip}\makeatother 
\IEEEpubid{\parbox{\columnwidth}{
		Workshop on Binary Analysis Research (BAR) 2025 \\
		28 February 2025, San Diego, CA, USA \\
		ISBN 979-8-9919276-4-2 \\ 
		https://dx.doi.org/10.14722/bar.2025.23004 \\   
		www.ndss-symposium.org
}
\hspace{\columnsep}\makebox[\columnwidth]{}}

\maketitle

\begin{abstract}
  This paper introduces \toolname{}, a tool for analyzing and
visualizing differences between two versions of a software binary. The
primary use case for \toolname{} is validating ``micropatches'': small
binary or assembly-level patches inserted into existing compiled
binaries. To perform this task, \toolname{} leverages the Python-based
\angrname{} symbolic execution framework.  Our tool analyzes the
output of symbolic execution to find end states for the pre- and
post-patched binaries that are \emph{compatible} (reachable from the
same input). The tool then compares compatible states for observable
differences in registers, memory, and side effects. To aid in
usability, \toolname{} comes with a web-based visual interface for
viewing comparison results. This interface provides a rich set of
operations for pruning, filtering, and exploring different types of
program data.


\end{abstract}


%
\IEEEpeerreviewmaketitle

\section{Introduction}
Much of today's infrastructure is built on a foundation of legacy
software; maintaining and securing this software is a critically
important task.  Patching legacy software must sometimes take place at
the binary level due to loss of source code, build
toolchain/environment ``bit rot,'' or limitations on the deployment
system (for example, bandwidth-limited systems in contested
environments). Under these conditions, software maintainers sometimes
deploy software \emph{micropatches}: minimal assembly-level changes
that fix a bug or add functionality.  Due to the low-level nature of
binary patches, it can be difficult to reason about their effects on
program behavior.

In theory, one could gain confidence that a patch has made all and
only the desired changes by using a variant of \emph{comparative
  symbolic execution}\footnote{Prior work sometimes refers to
  ``differential symbolic execution''
  \cite{differential_symbolic_execution} and ``relational symbolic
  execution'' \cite{farina}. Throughout this work, we use
  ``comparative symbolic execution'' as a generic term for this family
  of techniques.} (CSE) \cite{differential_symbolic_execution, farina}.
In other words, one could run the pre- and post-patched
programs on symbolic input in order to identify inputs that cause the
programs to behave differently or violate a relative correctness
specification.  However, two challenges limit CSE's suitability for
validating real-world binary patches. First, existing CSE techniques
target source code or idealized high-level languages. Second, CSE
results can be difficult to interpret. CSE typically produces a formal
description of the programs' semantic differences; this description
can be complex when the programs under analysis are large, when the
patch produces a large change in program behavior, or both.

This work presents \toolname{}, a tool that provides insight into the
effects of binary patches by identifying and visualizing semantic
differences between binary programs. The tool has two main components:
(1) a \emph{symbolic execution framework} for analyzing pairs of
binaries, and (2) a \emph{visualization engine} for displaying and
exploring CSE results. The \toolname{} approach to CSE involves
running two programs on symbolic input in order to identify pairs of
final machine states that are \emph{compatible}: reachable by the same
input. Differences in program behavior can then be characterized in
terms of differences between compatible states. One attractive feature
of this approach is that unlike some comparative analyses, \toolname{}
CSE does not require a correctness specification as input. This
feature is useful when the analyst does not know in advance how the
programs should differ. In such a case, the analyst can examine
compatible state pairs manually or check various specifications
against the pairs in a post hoc manner.

Because \toolname{} targets a scenario in which source code is
unavailable, it must be able to symbolically execute binary
programs. To achieve this goal, the tool builds upon the \angrname{}
\cite{ANGR} binary analysis platform.

In summary, this work makes the following contributions:
\begin{itemize}
\item{We present the \toolname{} comparative symbolic execution (CSE) framework, a novel adaptation of CSE to the binary domain.}
\item{We present the \toolname{} graphical interface for visualizing the results of CSE and for exploring the effects of binary patches on program behavior.}
\end{itemize}

\toolname{} is an open-source Python package. The tool can be
installed via the Python Package Index (PyPI) \cite{cozy-pypi}; its source code
and documentation are available on GitHub \cite{cozy-github}. 


\begin{figure*}[!h]
  \centering
  \begin{subfigure}{0.4\textwidth}
    \begin{minted}[linenos,numbersep=4pt,fontsize=\scriptsize,highlightlines={2-5}]{C}
void update(char *serialized) {
  // begin patch
  if (num_semicolons(serialized) > 2 ) {
    puts("bad serialization!"); exit(1); }
  // end patch
  char *command = strtok(serialized, ";");
  char *role = strtok(NULL, ";");
  char *data = strtok(NULL, "");
  if ((command is not "DELETE"|"STORE") ||
      (role is not "root"|"guest" )) {
    puts("bad input!"); exit(1); }
  if ((command is "DELETE") && (role is "root")) {
    delete(data);
  } else if (command is "STORE") {
    store(data);
  } else {
    puts("permission denied");
  }
  exit(0); }
int main(int argc, char **argv) {
  char *command = argv[1], role = argv[2],
    data = argv[3];
  int len = strlen(command) + strlen(role) +
    strlen(data) + 8;
  char *serialized = malloc(len * sizeof(char));
  sprintf(serialized, "%s;%s;%s", command,
          role, data);
  update(serialized); }
    \end{minted}
    \caption{Pseudo-C code for a simplified database front end. A user
      with root access is allowed to both store and delete data; a
      user with guest access is only allowed to store data. The
      original version of the program, which excludes lines 2--5, has
      a command injection vulnerability that enables a malformed
      command string to bypass the prohibition on guest
      deletions. Lines 2--5 are an overly restrictive patch that fixes
      the vulnerability but also rejects valid data payloads.}
    \label{fig:example-program}
  \end{subfigure}\quad\quad
  \begin{subfigure}{0.5\textwidth}
    \includegraphics[width=1.0\textwidth]{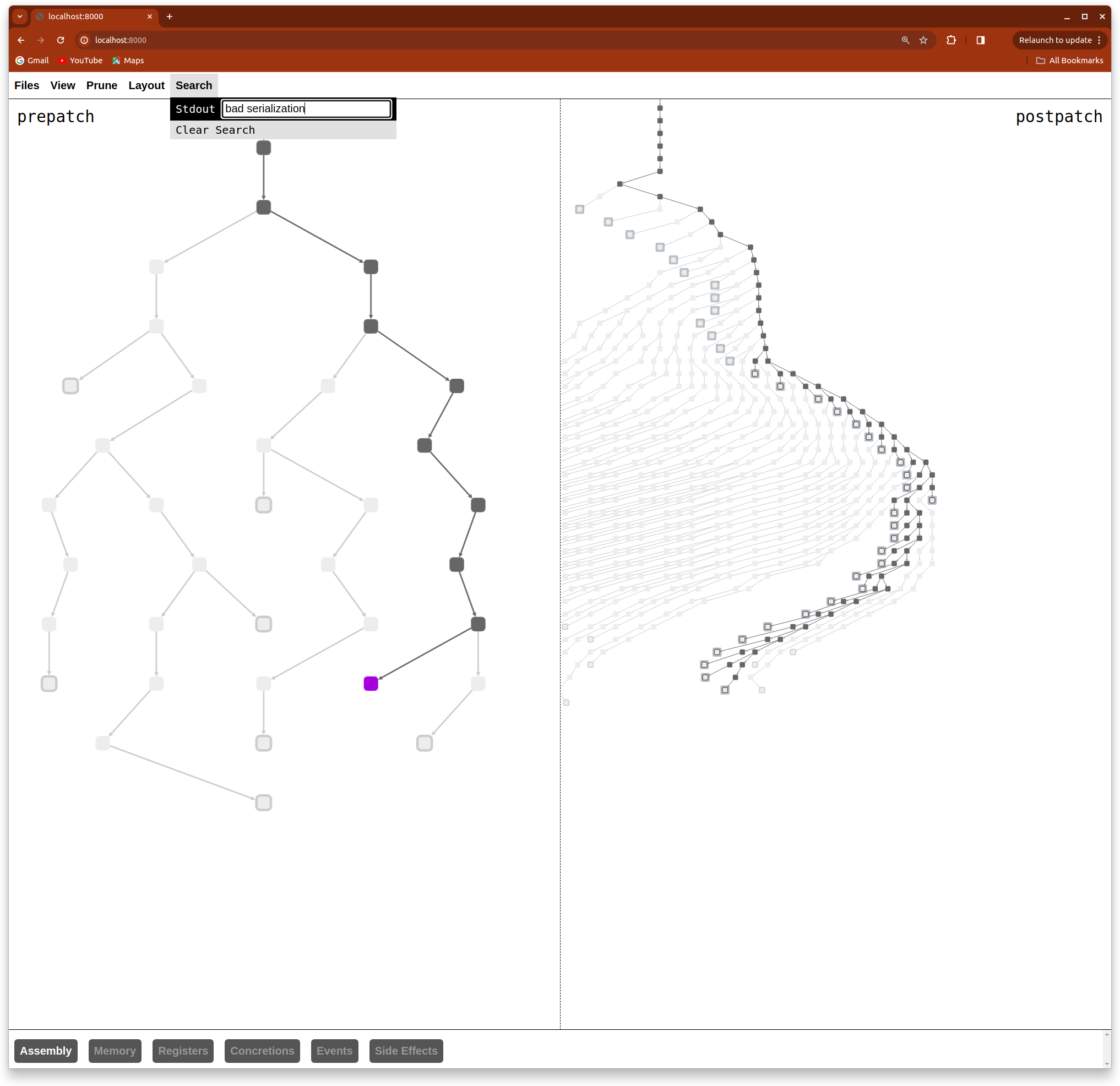}
    \caption{\toolname{} visual comparison of the pre- and
      post-patched versions of the Figure~\ref{fig:example-program}
      program. Trees on the left and right represent possible
      execution paths for the pre- and post-patched programs,
      respectively. The purple node on the left represents a violation
      of the assertion that a guest cannot delete data---i.e.,
      \toolname{} finds an input to the pre-patched binary that breaks
      the ``no guest deletions'' rule. The right pane shows paths
      through the post-patched binary that are triggered by the same
      input. All such paths, as well as some additional paths that are
      \emph{not} triggered by the input, have a square endpoint
      indicating that they print ``bad serialization!'' (the patch's
      error message). In other words, the patch rejects all
      vulnerability-triggering inputs, but it rejects some benign
      inputs as well.}
    \label{fig:example-ui}
  \end{subfigure}
  \caption{Program with patch (\ref{fig:example-program}) and
    \toolname{} visualization of CSE results for the pre- and
    post-patched program versions (\ref{fig:example-ui}).}
    \label{fig:example}
\end{figure*}

\section{Example}
\label{sec:example}

\sml{I started changing ``we'' to ``the user'' or ``the operator''
  throughout this section but gave up after a while. Let's decide
  which term we want to use and then edit accordingly.}

We introduce \toolname{} with an example that involves two attempts at
patching a vulnerable binary. \toolname{} helps the user discover that
while the first patch fixes the vulnerability, it also introduces
unintended behavior. The tool then confirms that the second patch
fixes the vulnerability without producing unintended behavior.

The example program, shown in
Figure~\ref{fig:example-program},\footnote{While \toolname{} operates
  directly on binaries, we present the program as pseudo-C source code
  for ease of understanding. An executable version of this example is
  available on the \toolname{} GitHub repo \cite{cozy-github}.} is a
simplified database server interface. The program's \texttt{update}
function takes a serialized string containing arguments
\texttt{command}, \texttt{role}, and \texttt{data} separated by
semicolons. The \texttt{command} argument must be ``STORE'' or
``DELETE'', the \texttt{role} argument must be ``root'' or ``guest'',
and \texttt{data} can be any string.  \sml{Should we use a code font
  for string literals?} \texttt{update} either (a) stores
\texttt{data} to the database, (b) deletes \texttt{data} from the
database, or (c) rejects the input as invalid, depending on the values
of \texttt{command} and \texttt{role}. A ``DELETE'' command is only
allowed when the role is ``root''; the check on line 12 enforces this
restriction. \sml{TODO: Check this line number after finalizing code
snippet.}The \texttt{main} function serializes the command line
arguments into a single semicolon-delimited string (line 17) and
passes the string to the \texttt{update} function.

The original binary, which corresponds to the pseudocode in
Figure~\ref{fig:example-program} minus highlighted lines 2--5, has a
command injection vulnerability: if the \texttt{role} argument is
``guest'' but the \texttt{command} argument is the string
``DELETE;root'', then the serialization-deserialization process
incorrectly allows a guest to delete data.

Lines 3--4 in Figure~\ref{fig:example-program} show an incorrect
patch, which reports an error if the serialized string contains more
than two semicolons. While this patch fixes the vulnerability, it is
overly restrictive because semicolons should be allowed in the
\texttt{data} payload argument, and the patch disallows such payloads.

To validate this change, the patch author runs \toolname{} on the pre-
and post-patched binaries. Doing so produces the visualization in
Figure~\ref{fig:example-ui}. The trees in the left and right panes
represent execution paths through the pre- and post-patched binaries,
respectively. The operator has used a \toolname{} feature to assert
that the \texttt{delete} function should never be called when the
\texttt{role} command line argument is ``guest'' (see Section
\ref{sec:directives} for details on assertions). The left (pre-patch)
pane includes a purple node that indicates an assertion violation; in
other words, \toolname{} identifies a path through the pre-patched
binary that corresponds to a command injection attack.

The user has clicked on the violation node in the left pane, which
highlights all \emph{compatible} paths in the right pane. Two paths
are \emph{compatible} when there is at least one concrete input that
causes execution to proceed down both paths (see Section \ref{sec:compatibility} for
a detailed discussion of compatibility). Additionally, we have
searched for paths in the right pane that print the string ``bad
serialization'' (the error message that the patch produces); all such
paths have a larger square endpoint.

We can immediately see that all paths compatible with the assertion
violation print ``bad serialization.'' In other words, the patch
rejects all inputs that would have triggered the vulnerability.
However, the right pane also shows several free-floating squares,
which are paths that print ``bad serialization'' but are incompatible
with the path to the assertion violation. Why is the patch rejecting
serialized input that would not have violated the assertion?

\begin{figure}[t]
  \centering
  \includegraphics[width=0.5\textwidth]{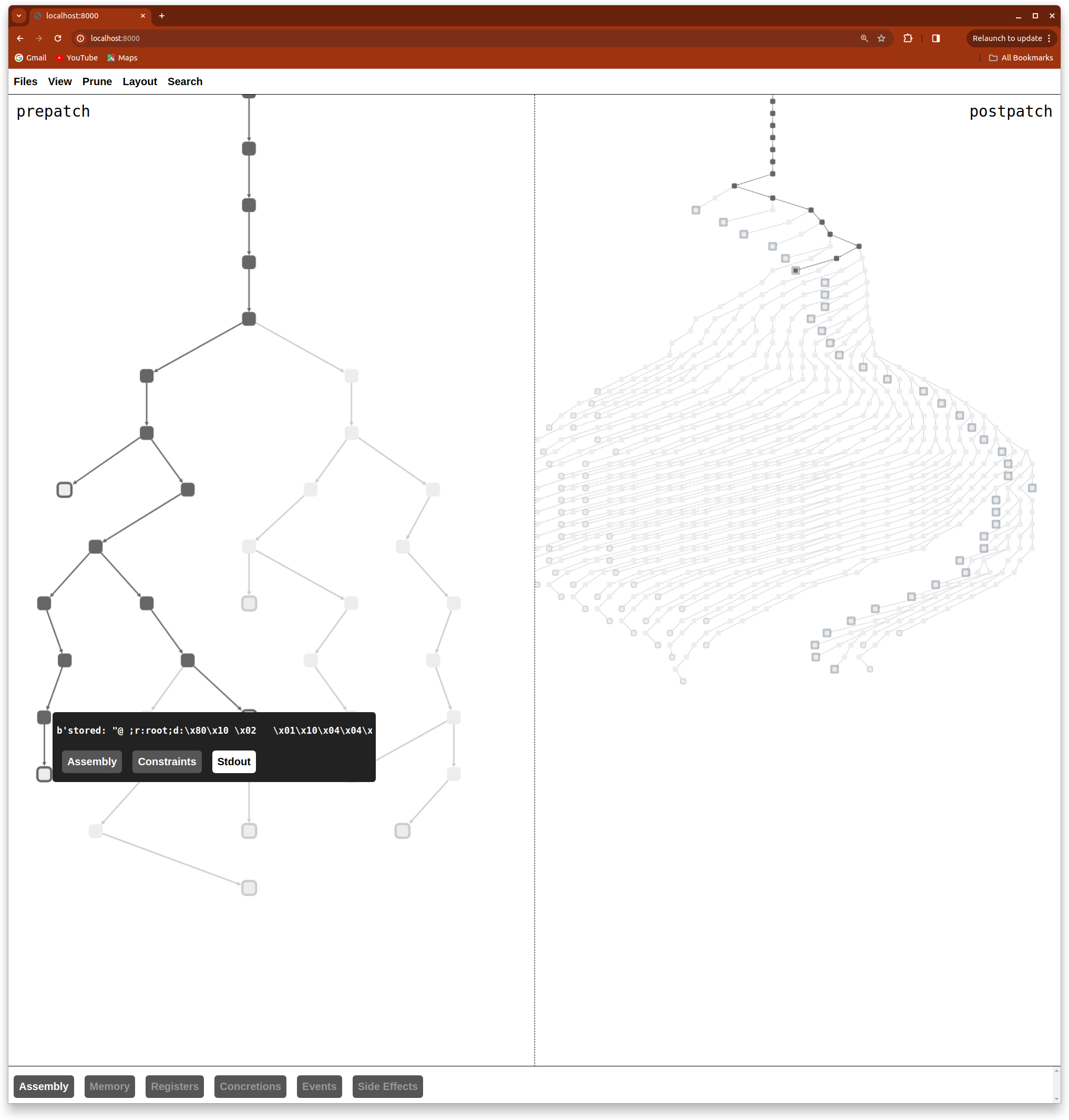}
  \caption{To understand why the first patch attempt rejects valid input, the user finds a violation-free path through the pre-patched binary that is compatible with a ``bad serialization'' path through the post-patched binary.}
  \label{fig:bad2}
\end{figure}

\begin{figure}[t]
  \centering
  \includegraphics[width=0.5\textwidth]{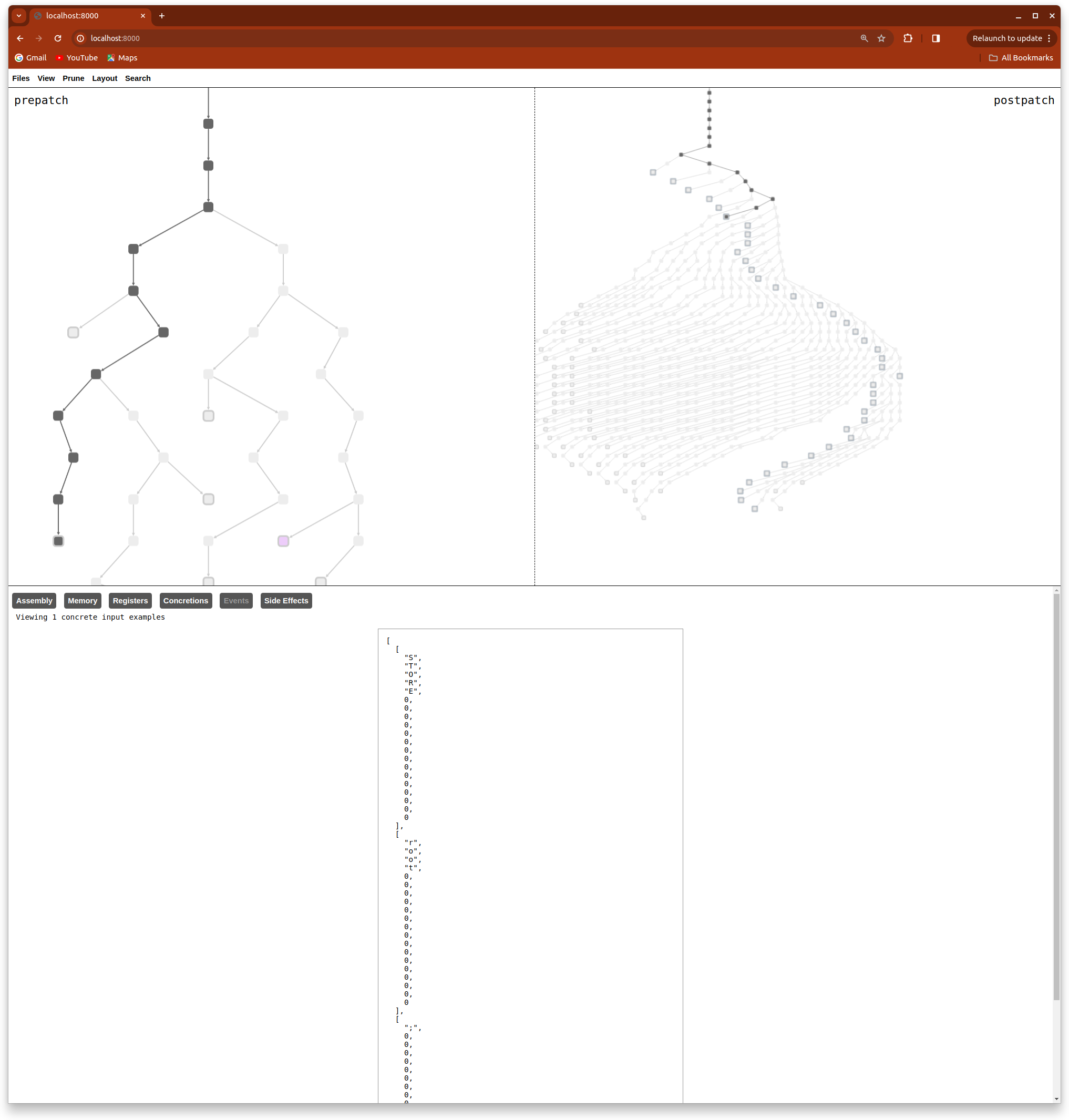}
  \caption{\toolname{} generates a concrete input that exercises the paths from Figure~\ref{fig:bad2}.  The input \texttt{command}=``STORE'', \texttt{role}=``ROOT'', \texttt{data}=``;'' is valid (\texttt{data} is allowed to contain semicolons), but the patch rejects it.}
  \label{fig:badpatchconcrete}
\end{figure}

To investigate further, we click one of the ``bad serialization''
matches in the right (post-patch) pane, and then hover over a
compatible endpoint in the pre-patch pane, as shown in
Figure~\ref{fig:bad2}. This sequence of actions corresponds to finding
a violation-free path through the pre-patched binary that the patch
would intercept. The standard output of the pre-patch endpoint shows
that this path involves a store operation. If we click that store
endpoint in the left (pre-patch) pane, we can ask \toolname{} for
concrete input(s) that triggers the corresponding
paths. \sml{It may not be clear that we've narrowed things down to one
  pre-patch path and one post-patch path; how about naming them (or
  their endpoints) $p$ and $p'$ or something?} As shown in
Figure~\ref{fig:badpatchconcrete}, \toolname{} synthesizes an input
that indeed has a semicolon in the \texttt{data} argument and that is
flagged as an error by the incorrect patch.

Finally, we replace the bad patch with a check in the \texttt{main}
function that the \texttt{command}
argument contains no semicolons, and we confirm that ``bad command''
errors arising from the new patch correspond to either (a) the
assertion condition, or (b) ``bad input'' conditions in the prepatched
binary.


\section{Comparative Analysis}
\label{sec:comparative_analysis}


\toolname{} uses symbolic execution to compare binary programs. The tool runs both programs on the same symbolic input until there are no remaining states to explore. Once symbolic execution is complete, \toolname{} pairs each terminal state in the pre-patched binary with each compatible terminal state in the post-patched binary. For each compatible pair, \toolname{} computes a diff of the pair's register contents, memory contents, and IO side effects. Once this process is complete, the user may either view the results in textual form or explore them via a graphical interface. In this section, we describe the program analysis that \toolname{} implements, and we outline the user's options for controlling and customizing that analysis.

\subsection{Setup}\label{sec:setup}

\begin{figure}[t]
  \includegraphics[width=0.5\textwidth]{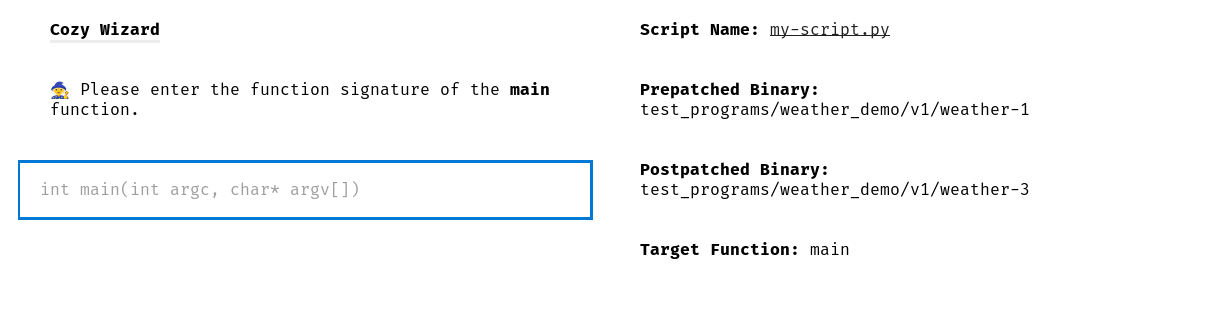}
  \caption{Interactive wizard that generates an application-specific \toolname{} harness based on user input. As shown here, one input to the wizard is the type signature of the function that will serve as the entry point for symbolic execution.}
  \label{fig:wizard}
\end{figure}        



\toolname{} typically runs in a \textit{harness}: a Python script that first configures various \toolname{} parameters and then invokes the tool on the target binaries. To streamline the process of creating an application-specific harness, \toolname{} provides an interactive wizard that asks the user a series of questions about how the tool should perform its analysis (see Figure~\ref{fig:wizard} for an example). The wizard generates a harness based on the user's responses. A typical harness performs the following steps:

\begin{enumerate}[left=0pt]
  \item Create \toolname{} projects for both binaries. A project is an object that acts as an interface between \toolname{} and a binary to analyze.
  \item Define any hooks that are needed to model hard-to-emulate functions. Hooks are common on embedded system targets where the callee function performs a side effect that cannot be modeled in the \angrname{} emulation environment.
  \item Create all symbolic variables that will be used during execution. Symbolic variables can represent function input as well as sources of nondeterminism. A common example of nondeterminism is when the program requires user input from stdin or over the network. For example, one may simulate the \texttt{getchar} function with a hook that returns a symbolic value.
  \item Define a \texttt{run} function that takes a project as input and symbolically executes its underlying binary, using the hooks defined in step \#2 along with any user-defined preconditions and initial memory values.
  \item Call the \texttt{run} function once on the pre-patched binary and once on the post-patched binary to produce run results containing lists of deadended states.
  \item Compare the run results to determine which state pairs are compatible, and then check each compatible pair for differences in registers, memory, and side effects.
  \item Launch a web browser window that shows a visualization of the comparison results.
\end{enumerate} 

\subsection{Compatible States}
\label{sec:compatibility}


Core to \toolname{}'s analysis and visualization is the notion of \textit{compatible} states.
We say that two terminal states $s$ and $s'$ are \textit{compatible} if there exists at least one concrete input that causes execution to terminate in state $s$ in the pre-patch execution, and in state $s'$ in the post-patch execution. We collect all compatible state pairs into the \textit{Compatible} relation. More formally:

\begin{definition}[Compatibility]
  \label{def:compat}
  \small
  \[\textit{Compatible} \triangleq \{ (s, s') \mid \textit{compatible}(s,s') \}\]
  \begin{equation*}
    \begin{aligned}
    \textit{compatible}(s, s') \triangleq \texttt{is\_sat}(s.\texttt{constraints}\,\wedge\\
    s'.\texttt{constraints})
    \end{aligned}
  \end{equation*}
\end{definition}

\noindent
where the notation \texttt{$s$.constraints} refers to the path constraints of terminal
state $s$.\footnote{Note that \angrname{} stores memory and register contents separately from path constraints; \toolname{} is built on top of \angrname{} and inherits this design choice.} 




\paragraph*{Unsat core optimization}
A na\"{i}ve way to compute the \textit{Compatible} set is to check all $n^2$ pairs of terminal states for joint satisfiability. \toolname{} implements a memoization-based optimization to enhance performance.
When \(s.\texttt{constraints} \wedge s'.\texttt{constraints}\) is unsatisfiable for a pair of states \((s,s')\), \toolname{} computes the \textit{unsat core} and caches it. The unsat core is the minimal set of clauses for which the conjunction is unsatisfiable. Later, when we want to know if a new pair $(s, s')$ is compatible, we first check if any previously discovered unsat core is a subset of the joint constraints $s.\texttt{constraints} \land s'.\texttt{constraints}$. If this check succeeds, then the joint constraints are immediately unsatisfiable, and we can skip the expensive call to \texttt{is\_sat}. Since most state pairs are incompatible in practice, the unsat core optimization drastically reduces the number of SMT solver queries.

\paragraph*{``No orphans'' property}
A desirable property of our analysis is the ``no orphans'' property; that is, 
every terminal state that the analysis reaches in one program should be compatible with at least one terminal state in the other program.
The ``no orphans'' property supports intuitive user interaction, such that whenever the user selects a path in one program, at least one corresponding path in the other program is highlighted.
The ``no orphans'' property holds for the symbolic execution strategies \toolname{} implements: complete execution (the variant described so far) and incomplete concolic execution (Section \ref{sec:concolic}). Each filter that the user can apply to the states through the \toolname{} interface (Section \ref{sec:visualization}) also preserves this property. We now give a proof for the complete execution case:


\begin{lemma}[No Orphans]
  \label{thm:no_orphans}
  After complete symbolic execution of two programs $P$ and $P'$, a terminal state $s_i$ from $P$ always has at least one compatible terminal state from $P'$.
\end{lemma}
\begin{proof}
  After complete exploration, the path conditions of the terminal state induce a disjoint complete partition over the set of possible inputs. Suppose that the input partition for the terminal states from $P$ is $\{X_0, X_1, ..., X_n\}$ and that the input partition for the terminal states from $P'$ is $\{Y_0, Y_1, ..., Y_m\}$.

  Because the inputs are the same for both programs, we have the following union condition:
  \[
    \bigcup_{i=0}^{n} X_{i} = \bigcup_{j=0}^{m} Y_{j}
  \]

  Assume that for state $s_i$ with corresponding non-empty input set $X_i$, the intersection with all $P'$ input sets $Y_j$ is empty. This is equivalent to saying that $s_i$ is an orphan.
  
  However, this would mean that there exists at least one concrete input $x \in X_i$ that cannot be found in the $P'$ input $\bigcup_{j=0}^{m} Y_{j}$. This contradicts the previous union condition which says that the input sets must be equal. Therefore, the state $s_i$ is not an orphan state.
\end{proof}

\subsection{IO Side Effects}
\label{ssec:io_side_effects}

In addition to comparing programs' final states, the \toolname{} user might wish to compare programs in terms of the \textit{side effects} that they produce. To enable this use case, \toolname{} has a subsystem for modeling IO side effects. Common examples of IO side effects that we have modeled in example programs include writing to stdout/stderr, writing to the network, and writing over a serial connection.

Modeling IO side effects with \toolname{} involves defining a \textit{hook} for a side effect-producing function that simulates the function's behavior. When symbolic execution reaches a call to a hooked function, \toolname{} runs the hook and stores the resulting side effect payload in the state. When a child state forks off from its parent, it obtains a copy of the parent's stored side effects. \toolname{} keeps track of IO side effects over different channels (stdout, network, etc.). When the user examines compatible states in the UI, \toolname{} visually aligns their side effects so that any differences are clear.

\subsection{Observational Differences}

Two compatible states with \textit{observational differences}---i.e., differences in their register values, memory values, or side effects---indicate the existence of an input that causes the two programs to behave differently. Because such differences may be of interest to users, \toolname{} checks each pair of compatible terminal states for equality of their registers, written memory, and IO side effects. Note that these state components may be a combination of concrete and symbolic values because \toolname{} runs programs on symbolic input.

For a compatible pair $(s, s')$, register contents $r$ in $s$ and register contents $r'$ in $s'$ are observationally different when the following condition holds:
\begin{equation}
  \label{reg_diff}
  \small
  \texttt{is\_sat(}s.\texttt{constraints} \land s'.\texttt{constraints} \land r \neq r')
\end{equation}
\noindent
\toolname{} constructs analogous conditions for memory writes and IO side effects, and it checks the conditions with an SMT solver. Because \toolname{} targets a micropatch scenario in which differences between programs are small, the tool is able to use several optimizations that reduce the number of SMT queries it must perform. Registers and memory values are often entirely concrete or syntactically identical, so they can be compared for equality without a solver query.

\toolname{} also employs a model-caching feature from \angrname{}'s built-in solver. When a formula like Condition \ref{reg_diff} is satisfiable, \toolname{} caches the model (concrete assignments that make the condition true). Later, when \toolname{} needs to determine whether a different formula is satisfiable, the tool checks whether any of the cached models satisfy the formula before it attempts to construct a fresh model.

\subsection{Directives}
\label{sec:directives}

\toolname{} supports several kinds of \textit{directives}, which are special hooks that run when execution reaches a specified program address. A directive can be thought of as a breakpoint that runs a snippet of user-provided code---for example, to debug symbolic execution or provide extra information to the execution engine. \toolname{} supports the following directives:

\begin{itemize}[left=0pt]
  \item \textbf{Breakpoint} pauses execution so that the program state can be inspected by user-provided Python code. When used in conjunction with a Python debugger, the simulation state can be inspected interactively.
  \item \textbf{Assume} attaches extra constraints to the program state when execution reaches a specified point.
  \item \textbf{Assert} by default operates like an assert in an ordinary programming language or testing environment. When \toolname{} performs a complete symbolic exploration, an assert can be used to ensure that for all possible inputs, the provided condition cannot be falsified. A common example of an assertion states that an array index stored in a register is in bounds before it is used in an array operation. Listing \ref{lst:assert_directive} gives an example of such an assertion.
  
  When symbolic execution encounters an assertion directive, it splits the current state into two child states: one in which the assertion is triggered, and one in which it is not. The state with the triggered assertion is stashed, and it is not executed further.

  \item \textbf{Postcondition} is a special type of assert that executes after the simulated function returns.

  \item \textbf{Virtual print} produces an IO side effect on the virtual print channel, which is useful for debugging an execution trace within the program. This technique is analogous to a symbolic version of \texttt{printf} debugging.
    
  \item \textbf{Error} is a directive that is triggered whenever the program reaches a specified address. When execution reaches an Error directive, \toolname{} stashes the current state; execution does not proceed further. This directive is useful for marking certain branches of the program as throwing an error.
\end{itemize}

\begin{listing}
\begin{minted}[fontsize=\footnotesize]{python}
def index_assertion(state: angr.SimState):
  index = state.regs.r2
  return (index.SGE(0) & index.SLT(BUFFER_SIZE))

session.add_directives(
  cozy.directive.Assert.from_fun_offset(
    project, "loop", 0x20,
    index_assertion, "index out of bounds"))
\end{minted}
\caption{Example of creating an assertion for an array bounds check. At instruction loop+0x20, we assert that the index (stored in register r2) must be in range. Note: SGE means ``signed greater or equal'' and SLT means ``signed less than.''}
\label{lst:assert_directive}
\end{listing}

\subsection{Concolic Exploration}
\label{sec:concolic}

By default, \toolname{} uses \angrname{}'s standard symbolic execution strategy of exploring non-terminal states in a breadth-first manner. As an alternative strategy, \toolname{} provides a variant of \textit{concolic execution}\cite{dart}. Concolic execution is desirable when the state space is large because it allows for incomplete exploration while still producing a set of final states that satisfy the ``no orphans'' property (Lemma \ref{thm:no_orphans}).

In the typical concolic execution scenario as presented in the literature \cite{concolic_testing}, the program first runs on a concrete input and generates an execution trace. Next, the program runs on symbolic input, which is forced to follow the concrete trace. After symbolic execution reaches a terminal state, a portion of the symbolic path condition is negated and a new concrete input is synthesized from this condition. This newly generated concrete input therefore exercises a different execution path.

\toolname{} achieves results similar to those of ordinary concolic execution, but it uses a different exploration process. When child states are generated from a parent, \toolname{} substitutes concrete inputs into their constraints; the tool then defers (halts exploration of) all children with constraints that evaluate to false. This approach obviates the need for separate concrete execution of the program; it fuses concrete and symbolic execution into a single process. This fusion decreases the engineering effort required to implement the concolic approach and integrate it with the existing complete exploration code.\sml{Okay?}

Once symbolic execution reaches a terminal state, \toolname{} uses one or both of the following heuristics to decide how to continue exploration:

\begin{enumerate}[left=0pt]
  \item \textbf{Termination Heuristic}: A termination heuristic determines whether \toolname{} should halt concolic execution. The default termination heuristic says that concolic execution should continue until the exploration of state space is complete. \toolname{} also enables the user to choose termination heuristics based on cyclomatic complexity and basic block code coverage metrics; these heuristics may lead to incomplete exploration. In addition, the user can define custom termination metrics.
  \item \textbf{Candidate Heuristic} If the termination heuristic says that exploration should continue, \toolname{} needs to decide which deferred state to explore next. Choosing a deferred state is equivalent to negating part of the path condition of a previous exploration.
  
  The ``trivial'' candidate heuristic simply chooses an arbitrary deferred state from the list of options. \toolname{} also provides a more complex n-gram branch coverage heuristic \cite{coverage_metrics_fuzzing} that attempts to choose the deferred state with the most unique basic block address history.
\end{enumerate}

Once the candidate heuristic chooses the next state to explore, \toolname{} generates a new concrete input from that state's path constraints. \toolname{} then feeds this concrete input into both programs under comparison by activating the appropriate deferred states (those with path conditions that are now satisfied). The program used to generate the concrete input alternates between the pre- and post-patch binaries to ensure that both versions of the function are being explored.

By feeding the same concrete input to both programs, \toolname{} ensures that no orphaned states will be generated. This invariant is important because it ensures that any terminal state a user selects in the \toolname{} UI is compatible with at least one state in the other program.



\section{Visualization}
\label{sec:visualization}

\begin{figure}[t]
  \begin{minipage}[t]{0.45\textwidth}
  \vspace{0pt}%
  \includegraphics[width=.99\textwidth]{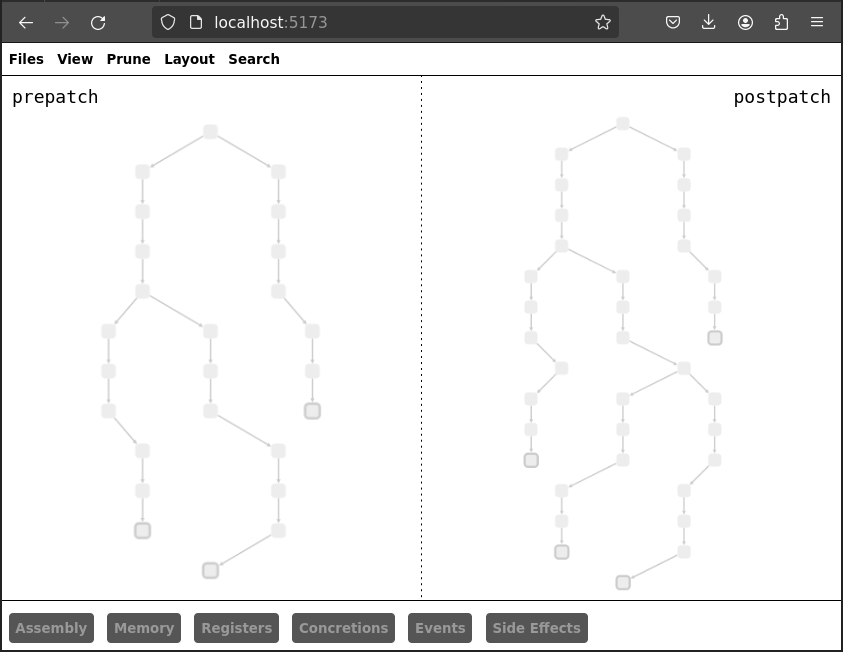}
  \end{minipage}\nobreak
  \begin{minipage}[t]{0.1\textwidth}
    \small
    \vspace{11pt}%
    \circled{1}

    \vspace{58pt}
    \circled{2}

    \vspace{73pt}
    \circled{3}
  \end{minipage}
  \caption{The \toolname{} GUI consists of (1) a menubar, (2) two panels displaying symbolic execution trees, and (3) a diff panel that enables the user to compare program branches across various dimensions.}
  \label{fig:ui_overview}
\end{figure}

The \toolname{} Graphical User Interface (GUI) is a simple web application. As shown in Figure~\ref{fig:ui_overview}, the GUI presents the user
with three main interfaces: (1) a menubar; (2) a pair of panels
displaying two symbolic execution trees; and (3) a ``diff panel,''
which presents detailed comparative information once the user selects
a pair of branches from the execution trees. In the remainder of
  this section, we describe the GUI's presentation of symbolic
  execution trees and its diff panel in more detail.

\subsection{Symbolic Execution Trees}

A symbolic execution tree depicts the results of symbolically executing a given program with \angrname. The root is the initial program state, an internal node is the program state after execution of a basic block, and an edge is a symbolic execution step.\sml{Added---accurate?}

When analyzing symbolic execution results, the user needs a way to cut
out extraneous noise. Typically, only a small subset of all of the
possible paths through a program are of genuine interest. The
\toolname{} GUI offers three main mechanisms for focusing on the
relevant parts of symbolic execution results: \textit{highlighting},
\textit{pruning}, and \textit{compression}.

Several types of program states that are likely to be significant are
automatically highlighted in the GUI. These include states that raised
errors during execution, states at which a syscall or
SimProc (modeled function) call occurred, states at which the program
exceeded user-specified boundaries on loop iteration,
and states at which a user-provided assert or postcondition
failed. Different colors indicate different categories of
potentially significant states. The color palate, and toggles to hide
or show each type of state, are available under the ``View'' menu
in the menubar.

Besides calling attention to relevant results, it can be helpful to
filter out irrelevant results. \toolname{}'s main mechanism for
filtering out irrelevancies is \emph{pruning}. Pruning works as
follows: \toolname{} prunes (hides) each branch unless it is
``interestingly related'' to a compatible branch in the facing tree,
where the user specifies (via the GUI) which relationships are
interesting. For example, the user can indicate that two branches are
interestingly related when their terminal states have different memory
contents; pruning will then leave only the branches that differ from
at least one compatible branch of the facing tree in terms of their
final memory contents. The relations that the GUI checks are
symmetric, so if a branch $b$ survives
pruning because it is partnered with a compatible branch $b'$, then
$b'$ will survive as well. Therefore,
pruning will never result in an orphaned branch.

Several pruning actions are available under the ``Prune'' menu. In
addition to memory differences, \toolname{} can check for differences
in register contents as well as stdout and stderr output. The tool can
also check whether at least one of two compatible branches ends with
an error state, and whether at least one branch produces stdout that
does not match a user-provided regular expression. In addition, the
user can apply multiple pruning relations simultaneously, which
results in pruning with the \emph{conjunction} of the selected
relations. We found that while it is possible in principle to apply
arbitrary Boolean combinations of prunings, this approach makes for an
excessively complicated UI. Hence, we restrict ourselves to the
simpler case of pure conjunction.

The final mechanism that \toolname{} provides for sorting through the
results of symbolic execution is \textit{compression}: merging
successive nodes that represent uninteresting or inevitable
computation steps. There are two available compression
levels: the user can (1) merge adjacent nodes that have identical
constraints and (2) merge every node that has a unique
child with that child node, eliminating all straight-line sequences of
symbolic states.\footnote{Symbolic execution can add a constraint without branching
  when, for example, the result of adding the negation of the
  constraint is unsatisfiable.}

Besides sorting through branches using highlighting, filtering, and
compression, the \toolname{} user must be able to extract information
from symbolic execution results. Within the GUI, there two primary
features that expose information about a
particular branch to a user. One of these
features---tooltips---offers simple at-a-glance information about the
nodes in a branch, taken in isolation. The other
feature---\toolname{}'s diff panel (Section \ref{diffpanel})---looks
at a branch in comparison with a compatible branch selected from the
facing tree.

A tooltip appears when the cursor hovers over a
node. Depending on the type of node, different kinds of information
are available. A tooltip displays the following information:
the assembly instructions provided by \angrname{}'s disassembler for
the given state; the representation of those instructions in
  VEX \cite{VEX} (the IR over which \angrname{} performs symbolic
  execution); the operative symbolic constraints;
and concrete examples of possible contents of stdout and
stderr. Special states---roughly those with special highlighting rules
as described above---expose additional information. For example,
error states expose error messages, and states that invoke SimProcs
give the name of the function being hooked as well as the library that
provides it.

To get a genuinely \emph{comparative} analysis, however, a user needs
to select two full branches as follows.  First, the user clicks on the
leaf of a candidate comparison branch, and that branch will be
highlighted, along with all compatible branches in the facing
tree. The user can then click on a compatible branch from the facing
tree and begin to use the diff panel, as described in the next
section.

\subsection{The Diff Panel}
\label{diffpanel}

The diff panel becomes available when the user selects a pair of
compatible branches for deeper analysis. The types of comparisons that
the diff panel supports can be grouped into three broad categories:
comparisons of \textit{event streams}, \textit{terminal states}, and
\textit{concrete inputs}.

\begin{figure}[t]
  \setlength{\fboxsep}{0pt}
  \fbox{\includegraphics[width=.475\textwidth]{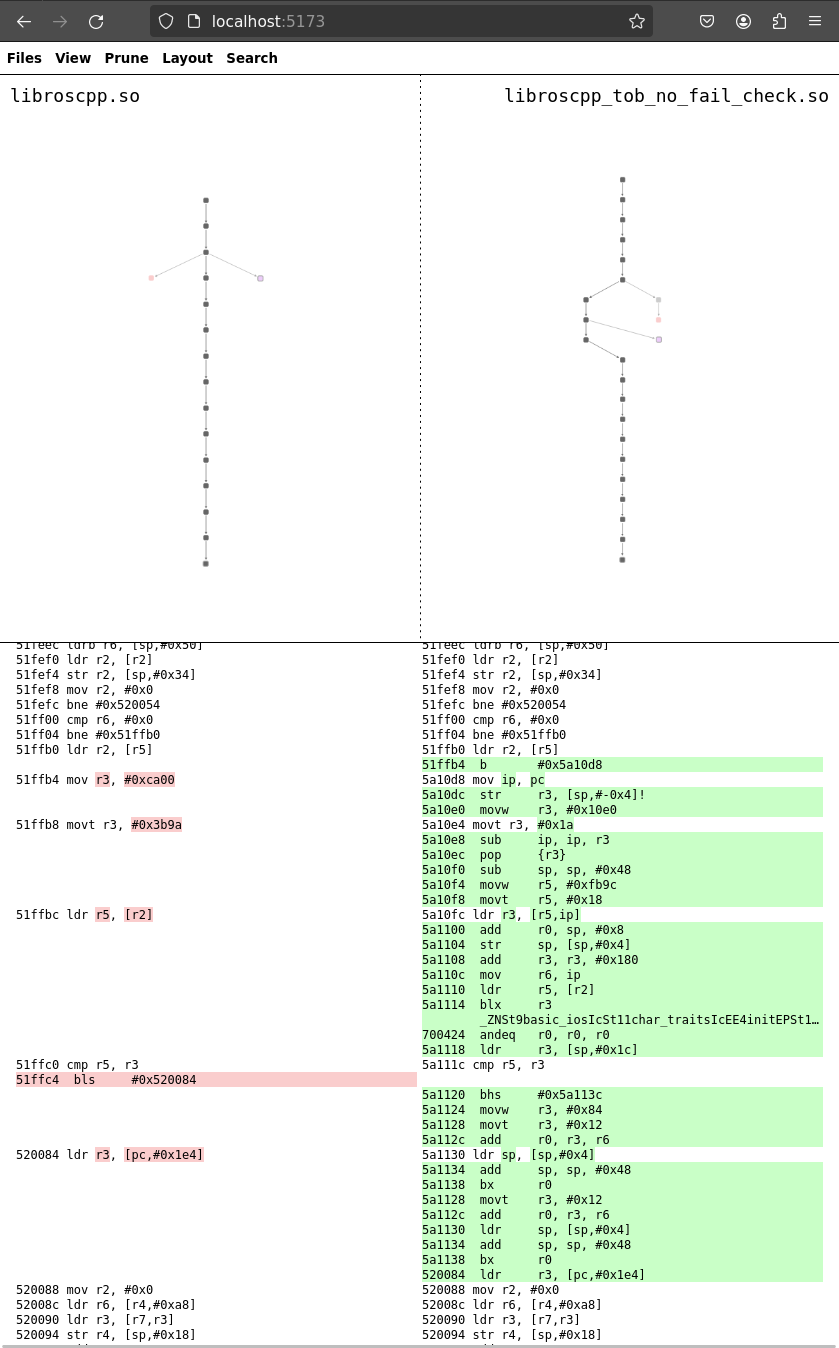}}
  \caption{Diff panel showing the assembly instructions in the original (left) and patched (right) versions of a program. Red and green highlighting represents deletion and insertion, respectively.}
  \label{fig:asm_diff}
\end{figure}

The sequence of nodes along a symbolic execution path corresponds to
several different kinds of event streams: the stream of assembly
instructions executed, the stream of read and write operations on
memory and registers, and the stream of modeled IO
effects. \toolname{} compares these types of event streams using a
familiar git-style line diff. An example of an assembly stream
comparison appears in Figure \ref{fig:asm_diff}, where it is possible
to see the exact region where program execution passes through a small
patch applied to a shared object file.

For each type of event stream comparison, when the user
mouses over an event, the UI highlights the tree node that corresponds
to that event. This behavior enables the user to intuitively
connect the contents of the tree-view to the contents of the event
stream. In some cases, the event stream also contextually exposes
other types of information. For example, the stream of assembly
instructions can provide the location in the original
source that corresponds to a given line of assembly, if this information
is recoverable from DWARF debug information in the binaries that
\toolname{} has analyzed.

In addition to event stream comparisons, \toolname{} supports
comparisons of terminal states.  For example, \toolname{} can compare
the final memory contents of two compatible branches. This process may
involve comparing symbolic values, since terminal states can contain
symbolic values. In such a case, \toolname{} checks whether the
symbolic values in the states are logically equivalent. If they are,
\toolname{} reports this fact, and if they are not, \toolname{}
generates some \emph{concretions} that illustrate a possible scenario
in which the terminal states differ in spite of an identical initial
state.

Finally, the diff panel can generate concrete inputs that
exercise compatible branches under comparison.
Compatibility guarantees the existence of an
input that produces the two sequences of behaviors that the branches
represent. The concretion view in the GUI's diff panel
displays example inputs that are shared between the two compatible
paths. This feature, in combination with
\toolname{}'s pruning functionality, make it possible to recover
specific inputs that generate execution paths of interest, especially
paths where behavior differs interestingly between the two binaries
being compared.

Compatibility does not guarantee that \textit{every} input that produces the behavior
associated with the first branch also produces the behavior associated with the
second branch, or vice versa. In cases where there are inputs that will produce
the behavior of the first branch, but not the second (or vice versa), \toolname{}
also makes these inputs available, and in cases where no such inputs exist,
\toolname{} makes it clear that one of the two branches ``refines'' the other,
or that the two branches are ``equivalent,'' in the sense that they represent
behaviors that are produced by exactly the same set of inputs.


\section{Evaluation}
\label{sec:evaluation}
\begin{figure*}[t]
    \centering
    \begin{tabular}{@{}l @{\hskip -0.5cm} S[table-format=2.0] @{\hskip 1cm} S[table-format=2.4] S[table-format=2.4] S[table-format=1.4]@{}}
        \toprule
        \textbf{Function Name} & {\textbf{\# Terminal States}} & \multicolumn{2}{c}{\hskip -0.2cm \textbf{Symbolic Execution Time (s)}} & {\textbf{Comparison Time (s)}} \\ 
        \cmidrule(r){3-4}
         &  & \multicolumn{1}{c}{Original} & \multicolumn{1}{c}{Instrumented} &  \\ \midrule
        \texttt{base64\_decode\_alloc\_ctx} & 31 & 16.0683 & 41.1273 & 2.0919 \\
        \texttt{base64\_decode\_ctx} & 31 & 15.1894 & 38.2189 & 2.0648 \\
        \texttt{base64\_decode\_ctx\_init} & 1 & 0.0108 & 0.0685 & 0.0451 \\
        \texttt{base64\_encode\_alloc} & 17 & 6.8573 & 10.5143 & 3.9269 \\
        \texttt{base64\_encode} & 17 & 7.6123 & 12.4908 & 4.6590 \\
        \texttt{clone\_quoting\_options} & 1 & 0.1203 & 0.1525 & 0.0461 \\
        \texttt{close\_stdout} & 1 & 1.4348 & 5.3738 & 0.0699 \\
        \texttt{close\_stdout\_set\_file\_name} & 1 & 0.0098 & 0.0656 & 0.0469 \\
        \texttt{close\_stdout\_set\_ignore\_EPIPE} & 1 & 0.0092 & 0.0629 & 0.0442 \\
        \texttt{close\_stream} & 58 & 6.2557 & 16.2690 & 4.0236 \\
        \texttt{decode\_4} & 29 & 4.3319 & 6.7472 & 1.5532 \\
        \texttt{deregister\_tm\_clones} & 1 & 0.0125 & 0.0529 & 0.0453 \\
        \texttt{fadvise} & 2 & 0.1976 & 0.1738 & 0.1010 \\
        \texttt{get\_quoting\_style} & 1 & 0.1713 & 0.1030 & 0.0506 \\ 
        \texttt{isbase64} & 1 & 0.1334 & 0.0849 & 0.0401 \\ 
        \bottomrule
    \end{tabular}
    \caption{Binary functions from a Linux \texttt{base64} utility, numbers of terminal states that \toolname{} symbolic execution finds for them, and running times for \toolname{} symbolic execution and verification. For each function, the original and instrumented versions have the same number of terminal states because the instrumentation code is branchless. A comparison time is the total time spent comparing all pairs of terminal states drawn from an original and an instrumented function.}
    \label{fig:eval}
\end{figure*}

We evaluated \toolname{} by measuring the tool's execution time as it
symbolically executed pairs of binary functions and checked a relative
correctness property over each pair. The evaluation goals were as
follows: (1) to observe \toolname{}'s execution speed on widely used
binary functions; (2) to demonstrate \toolname{}'s ability to verify a
desirable correctness property; and (3) to produce a set of
\toolname{} test harnesses that can serve as a reference point for
other users of the tool.
\subsection{Data Set}
Each instance in the data set is a pair of binary functions:
\begin{enumerate}
\item{A function $f$ taken from a Linux \texttt{base64} binary}
\item{A modified version of $f$ instrumented with code that supports coverage-guided fuzz testing}
\end{enumerate}

To create the data set, we used the RetroWrite binary rewriting tool
\cite{dinesh} to instrument the \texttt{base64} binary with code that
supports integration with the American Fuzzy Lop (AFL) fuzzer
\cite{zalewski}. We then selected 15 functions from the original
binary and paired them with their instrumented versions from the
modified binary.

\subsection{Correctness Property and Experimental Setup}
Because the instrumentation only exists to support fuzzing, a function
from the original binary should have the same observable behavior as
its instrumented counterpart. We use \toolname{} to verify this
property as follows. First, \toolname{} symbolically executes both
functions and computes the set of compatible state pairs. Second, for
each pair, \toolname{} checks an assertion that the states agree in
terms of their register and memory contents. If \toolname{} can
falsify this assertion, then there exists an input that causes the two
functions to behave differently, and verification fails.

The precise formulation of state agreement depends on a function's
return type. For example, if a function returns a 64-bit integer, then
two compatible states hold equal return values when the full contents
of their \texttt{RAX} registers are equal (\texttt{RAX} is the 64-bit
return register for the x86-64 ISA). However, in the case of a
function that returns a 32-bit integer, only the lower 32 bits of
\texttt{RAX} (i.e., register \texttt{EAX}) must be equal across the
states---the higher-order bits of \texttt{RAX} are allowed to
differ. Parameter types place similar constraints on the
functions' symbolic input. For these reasons, each data instance
requires a custom test harness that captures function-specific
behavior. These harnesses, along with our full evaluation
framework, are included in the public \toolname{} repository
\cite{cozy-github}.

\subsection{Results}
Using the process described above, we checked each function pair in
the data set for equivalent observable behavior. The evaluation took
place on a machine running Ubuntu 20.04 with an Intel i9-12900H
processor and 64 GB of RAM. The results appear in
Figure~\ref{fig:eval}. The table shows symbolic execution time for the
original and modified binaries, as well as comparison time, which
includes time spent computing compatible states and comparing register
and memory contents. \toolname{} verifies that the instrumentation
code leaves each function's observable behavior unaffected.

\section{Related Work}
\label{sec:related_work}
Computing differences between programs has a long history in the literature. Unlike the symbolic execution discussed here, the majority of previous tools operate on the textual or abstract syntax tree (AST) level \cite{gumtree, changedistilling,changedetection}, and do not attempt any actual simulation of the programs under analysis.

The \texttt{diff} utility\cite{unix_diff} distributed with Unix based operating systems is one example of an early comparison program. \texttt{diff} reports differences in lines, and performs a longest common subsequence computation to attempt to align two text files. The \texttt{diff} utility is generic, in the sense that it will function over any programs that can be represented in text files. However this approach, because it does not understand the semantics, cannot be used to provide a rich understanding of program behaviour.

\toolname{} does utilize a textual diff over the assembly trace (see Figure \ref{fig:asm_diff}) of a program in the visualization interface. When two terminal states are selected, the assembly pane will give a linear list of instructions executed for that trace, in the format of color-coded line based diff.



The most relevant prior work to our approach is that of Person et al. in their paper on ``Differential Symbolic Execution'' \cite{differential_symbolic_execution}. Our approach differs in a number of key ways. First, we analyze binary programs, whereas Person's approach analyzes high-level Java programs. Second, the method by which we check for pair compatibility and report deltas differs. In Person's computation of the partition effect delta, path conditions are checked for strict equivalence using an ``if and only if.'' This approach may detect inconsequential changes in control flow. Our approach is only concerned with observational differences---differences in registers, memory, and IO side effects after execution. It ignores differences at intermediate execution points that Person's tool would flag. Finally, our analysis of final register, memory and IO side effect content is more fine-grained than Person's approach, which has enabled us to create a novel visualization interface.

Person additionally discusses symbolic summary, which we do not utilize in our execution model. \textit{Symbolic summaries} may be used to summarize the effects of common blocks of code. Additionally, \textit{abstract summaries} may be used to skip execution of code that the two programs under comparison have in common. For example, a common code block $B$ when fed identical inputs (registers and memory) will result in the two programs reaching an identical ending state, regardless of the actual execution that occurs within $B$.

C standard library hooks are one location where symbolic summaries are currently used in \toolname{}. These hooks intercept calls to standard library functions, and perform the equivalent computation via a Python callback. The hooks are meant to simplify hard to execute standard library functions, typically resulting in far fewer child states.

Abstract symbolic summaries, while providing interesting benefits, do suffer from several drawbacks that makes them infeasible to use in \toolname{}. Due to their black box nature, abstract symbolic summaries do not allow for fine-grained analysis of register and memory contents in terminal states. Additionally, abstract symbolic summaries, since they are essentially computation that is skipped, do not allow for generation of concrete example inputs that lead to selected terminal states. In our experience with the micropatching process, generating concrete example inputs is essential for aiding in understanding program behaviour.

\textit{Shadow symbolic execution} is another body of work \cite{Palikareva2016-rd, Noller2018-up, Kuchta2018-ze} that functions on principles similar to \toolname{}. In shadow symbolic execution, an original and patched program are symbolically executed in lockstep until divergence is reached. Divergent program points are used to generate new test cases that exercise the impact of the patch. Divergence must be manually annotated by constructing a combined original and patched program via a special \texttt{change()} macro.

\toolname{} differs in several key ways from shadow symbolic execution. \toolname{} executes the original and patched binary in two separate symbolic execution runs, removing the need for manual \texttt{change()} annotations. Additionally, \toolname{} operates on binary programs, whereas the literature on shadow symbolic execution has focused on Java, C, and C++ programs.




\section{Discussion}
\label{sec:discussion}
As part of the DARPA Assured Micropatching (AMP) program, we tested
\toolname{} on a variety of third-party challenge problems. For
example, we used \toolname{} to (1) examine a proposed micropatch for
the Army MRZR platform; (2) identify a shortcoming in the initial
patch; and (3) show that all execution paths are correctly handled
with an improved patch \cite{mrzr}. We have additionally created a
variety of example programs designed to exercise different portions of
the tool. In this section, we discuss our observations of the
micropatching process and how \toolname{} performs in the overall
workflow.

The primary challenge of understanding micropatch behavior is making sense of the large volume of information that \toolname{} generates. For all but the simplest programs, the textual report \toolname{} generates is too cumbersome to understand. This fact led to the creation of the interactive visualization interface.


Direct examination of the symbolic values attached as state constraints, or stored in registers or memory is generally unhelpful. These symbolic expressions are typically large and too complex to be easily understood with manual inspection. \toolname{}'s ability to generate example concrete inputs, for a pair of compatible states, has proven both intuitive and useful.


States with assertion failures are flagged with a purple color, making them easily visible in the tree view. One common workflow is to check that all assertions triggered in the prepatched program are not triggered in the postpatch program. Prepatched assertion failures should be compatible with postpatch states that jump to micropatch code. By exploring various execution traces, concrete examples, and comparisons, the operator can achieve a high degree of assurance that the micropatch is behaving exactly as intended.

The skills required to use \toolname{} overlap with those needed to use \angrname{}. A rough understanding of assembly code is required to attach assertions at certain program points. The initial effort to apply \toolname{} to two versions of an application is outlined in Section~\ref{sec:setup}. The top-level arguments must be constructed, which requires knowledge of the argument types and their memory layouts. One can obtain this information from original source code or from a reverse engineering tool like Ghidra\cite{ghidra}.\sml{Source code won't give you memory layout, right?}

Since \toolname{} uses symbolic execution as its base analysis, it inherits the challenges of that technique: path explosion, nontermination, and costly SMT queries. To mitigate the path explosion problem, we have implemented joint concolic execution (Section \ref{sec:concolic}). The concolic execution we have implemented may be used for incomplete exploration while preserving terminal state compatibility. The difficulty of generating ``interesting'' concrete inputs is still a weakness of this approach. Although the heuristics attempt to explore deferred execution states that have unique basic block histories, we cannot know what concrete inputs will lead to interesting \textit{future} states.

Non-termination presents another problem for symbolic execution. It is obviously difficult, in general, to detect non-termination. In some programs, non-termination is a feature; for example, in event-handling loops. To deal with non-termination, we allow the user to place an upper bound on the number of times a loop executes.
As a simple mechanism to avoid nontermination, \toolname{} uses \angrname{}'s \texttt{LocalLoopSeer} exploration technique, which detects loops by recording the history of execution. If the upper bound on instruction iteration count is reached, we halt execution of that state and stash it. In the visualization, the spinning state can be seen as a downwards facing arrow.

In this paper, we haven't yet touched on the creation of formal specifications for intended patch behavior. Our initial work on the DARPA AMP program focused on this area and heavily utilized the CBAT tool \cite{cbat}. A formal specification boiled down to creating an SMT formula with an if-then-else (ITE) at the top level. The condition of the ITE determined when the patch changed program behavior, the true branch specified how the patch changed behavior, and the false branch specified that memory and registers must be identical in all other circumstances.

Tool operators had several complaints about creating these formal specifications: (1) the specifications were difficult to write, requiring the construction of complex SMT formulas; and (2) writing a formal specification was similar to writing the patch in the first place, so there were complaints about having to do the same work twice. Based on feedback from these third-party operators, we determined that an interactive, visualization-based approach would be more helpful.

The feedback loop created by the \toolname{} tool is, in essence, an interactive way to explore the formal specification space. It is possible to use \toolname{} to check directly that a patch changes behavior only in some specified way. This kind of formal verification is accomplished by writing a function that takes in a compatible state pair and returns an assertion condition. If \toolname{} can falsify the assertion for any compatible state pair, then verification fails.



\section{Conclusion}
\label{sec:conclusion}
In this paper we have presented \toolname{}, a Python-based framework
built on top of \angrname{} that uses symbolic execution to detect
observable differences in binary programs. The \toolname{} project is
designed to analyze micropatches, which are small binary or
assembly-patches inserted into existing legacy programs. By using
\toolname{}'s novel visualization interface, the tool's operator can
gain confidence that a given micropatch has its intended
effect. Operators who already have experience with the \angrname{}
symbolic execution framework will find it easy to get started with
\toolname{}. We hope that operators will find \toolname{} useful as
part of the verification step of the micropatch development process.




\bibliographystyle{IEEEtran}
\IEEEtriggeratref{11}
\bibliography{paper}



%

\end{document}